\documentclass[12pt]{article}
\begin{document}
\begin{center}
{\bf Vacuum  Energy Problem, Fundamental Length and Deformed
Quantum Field Theory}\\ \vspace{5mm} A.E.Shalyt-Margolin
\\ \vspace{5mm}
\textit{National Center of Particles and High Energy Physics,
Bogdanovich Str. 153, Minsk 220040, Belarus}
\end{center}
\vspace{0.5cm}
\begin{abstract}
The cosmological constant (vacuum energy) problem is analyzed
within the scope of quantum theories with UV-cut-off or
fundamental length. Various cases associated with the appearance
of the latter are considered both using the Generalized
Uncertainty Relations and the deformed density matrix,previously
introduced in the author's works. The use of the deformed density
matrix is examined in detail. It is demonstrated that, provided
the Fischler-Susskind cosmic holographic conjecture is valid, the
Vacuum Energy Density takes a value close to the experimental one.
The arguments supporting the validity of this conjecture are given
on the basis of the recently obtained results on Gravitational
Holography.
\end{abstract}
\section{Introduction}
The problem of vacuum energy is one of the key problems in a
modern theoretical physics. This problem has attracted the
attention of researchers fairly recently with the understanding
that a cosmological constant determining the vacuum energy density
is still nonzero despite its smallness.  As is known, the
cosmological constant $\Lambda$ has been first introduced in the
works of A.Einstein \cite{Einst} who has used it as a
antigravitational  term to obtain solutions for the equations of a
General Relativity (GR) in the stationary case. However, when
A.Friedmann has found the solutions for GR in case of expanding
Universe \cite{Fried}and E.Hubble has derived an extension of the
latter, A.Einstein refused from the cosmological constant calling
it "his greatest error". Specifically, in his letter to H.Weyl in
1923, Einstein's comments on the discovery of expanding Universe
were something like \cite{Pais}"If a quasistatic world is
nonexistent the cosmological term should be dropped!" But
$\Lambda$ has survived due to the above-mentioned reasons. The
principal problem of the cosmological constant resides in the fact
that its experimental value is smaller by a factor of $10^{123}$
than that derived using a Quantum Field Theory (QFT).
\\And the theories actively developed at the present time
(e.g., superstring theory, loop quantum gravity, etc.) offer a
modified quantum theory including, in particular, the fundamental
length at Planck's scale. The estimates of $\Lambda$ obtained on
the basis of these theories may be greatly differing from the
initial ones derived from the standard QFT.
\\ This paper presents a brief discussion of the relationship
between the appearance of the Fundamental Length and Vacuum Energy
Density Problem or Cosmological Constant Problem, from different
viewpoints and with the use of various approaches. Note that the
Vacuum Energy Density Problem is the main part of a more general
problem -- Dark Energy Problem.

\section{Vacuum Energy and Generalized Uncertainty Relations}
As a model, the fundamental length was introduced in the works
devoted to QFT about fifty years ago. It should be noted that such
a length was not necessarily associated with the Planck scales.
But due to the development of a string theory, Heisenberg's
uncertainty relations were modified in the 80-ies of the XX
century. These new relations were called the Generalized
Uncertainty Relations (GUR) \cite{Shen1}:
\begin{eqnarray}\label{xp}
\Delta x\geq \frac{1}{\Delta p}+\lambda (\Delta p),
\end{eqnarray}
that implies a minimal length $2\sqrt{\lambda}$, where $\lambda$
is on the order of the Planck area $l_p^2\sim G$.
\\GUR enables one to estimate the value of $\Lambda$
for several cases, in particular for the Schwarzschild-de Sitter
space-time ({\bf SdS}) with the geometry \cite{Shen1}
\begin{eqnarray}
ds^2=-\left(1-\frac{2M}{r}-\frac{1}{3}\Lambda
r^2\right)dt^2+\left(1-\frac{2M}{r}-\frac{1}{3}\Lambda
r^2\right)^{-1}dr^2+r^2d\Omega.
\end{eqnarray}
These spaces are in many respects similar to Schwartzshild's black
holes. To illustrate, they possess a horizon, and the notion of
temperature may be introduced for them like for the case of a
black hole when the temperature $T$ is still defined in terms of
the surface gravity of horizons. Using the formula for $T$ and
entropy  $S$ in ({\bf SdS})derived in \cite{Shen1}, and also the
correction to these values, with the use of GUR the authors have
succeeded in approaching the real estimate of the cosmological
constant $\Lambda$ in the case under study
\begin{eqnarray}
\frac{\Lambda}{m_p^2}\sim 10^{-120}.
\end{eqnarray}
However, the importance of this approach is somewhat lowered as it
is applicable only to the case studied or to the case of de Sitter
space-time.
\\ A more general approach with the use of GUR has been
first proposed in \cite{Ahl}. In so doing GUR were involved into a
more general context: stable deformation of Poincar\'e-Heisenberg
algebra. Here "stable" means deformation immunity, with the
retained structural constants of the associated Lie algebra. In
the previous works \cite{Vilela} and \cite{Chrys} it has been
shown that only the Lie algebras immune to infinitesimal
deformations may be significant in physics. In \cite{Ahl} a
deformation of of Poincar\'e-Heisenberg algebra is constructed
stabilized Poincar\'e-Heisenberg algebra (\texttt{SPHA}). Apart
from GUR, it carries three additional parameters: the length scale
pertaining to the Planck/unification scale, the second length
scale associated
 with cosmos, and a new dimensionless constant.
\\Note that such an extension of Poincar\'e-Heisenberg
algebra is always leading to the noncommutative space-time, and
the coordinate operators $X_{\mu}$ and $X_{\nu}$ for different
$\mu$ and $\nu$ do not commute. Specifically, in the case
considered in \cite{Chrys},\cite{Ahl} the appropriate commutative
relation takes the form
\begin{eqnarray}
\left[X_\mu,X_\nu\right] =  i {\ell^2_U} J_{\mu\nu},
\end{eqnarray}
where $J_{\mu\nu}$ are generators of the rotation group and $\ell_U
=\gamma \ell_P$. $\gamma$ is a new constant: $10^{-17} \le \gamma
\le 1 $. $\ell_P$ as usual is a Planck scale.
\\ By the approach developed in \cite{Ahl} the inclusion
of the commutation relations (\texttt{SPHA}) may result in
modification of an expression for the vacuum energy density, with
its value approaching the real one. This is demonstrated taking as
an illustration zero point energy for a simple harmonic oscillator
\cite{Vilela}. It is assumed that the proposed method may be
extended to give estimates for the effective value of the
cosmological constant $\Lambda$.
\section{Deformed Density Matrix, Holographic Principle
 and Vacuum Energy}
As it has been repeatedly demonstrated earlier, a Quantum
Mechanics  of the Early Universe (Plank Scale) is a Quantum
Mechanics with the Fundamental Length
(QMFL)\cite{Gar1}--\cite{Magg1}. In the works
\cite{shalyt1}--\cite{shalyt10} an approach to the construction of
QMFL has been developed with the help of the deformed density
matrix, the density matrix deformation $\rho(\alpha)$ in QMFL
being a starting object called the density pro-matrix and
deformation parameter (additional parameter)
$\alpha=l_{min}^{2}/x^{2}$, where $x$ is the measuring scale and
$l_{min}\sim l_{p}$.
\\As has been indicated in \cite{shalyt1}--\cite{shalyt10}
deformation parameter $\alpha$ is varying within the limits
$0<\alpha\leq1/4$, moreover $\lim\limits_{\alpha\rightarrow
0}\rho(\alpha)=\rho,$ where  $\rho$ being the density matrix in
the well-known Quantum Mechanics (QM). The explicit form of the
above-mentioned deformation gives an  exponential ansatz:
\begin{equation}\label{U26S}
\rho^{*}(\alpha)=\sum_{i}\omega_{i} exp(-\alpha)|i><i|,
\end{equation}
where all $\omega_{i}>0$ are independent of $\alpha$ and their sum
is equal to 1.
\\The correspondent deformed quantum field theory is defined at
the  non-uniform lattice in hypercube $I_{1/4}^{4}$ with the side
1/4 in length and edge of $I_{1/4}=(0;1/4]$
\cite{shalyt6},\cite{shalyt7}. It is designated as $QFT^{\alpha}$.
 All the variables associated with the considered $\alpha$ -
deformed quantum field theory  are hereinafter denoted by the
upper index $^{\alpha}$.
\\Then $QFT^{\alpha}$  will be compatible with the holographic
principle, i.e. with the holographic entropy bound  derived in the
earlier works \cite{Hooft1},\cite{Hsu}.
\\As follows from the
holographic principle, the maximum entropy that can be stored
within a bounded region $\Re$ in $3$-space must be proportional to
the value $A(\Re)^{3/4}$, where $A(\Re)$ is the surface area of
$\Re$. Of course, this is associated with the case when the region
$\Re$ is not an inner part of a particular black hole. Provided a
physical system contained in $\Re$ is not bounded by the condition
of stability to the gravitational collapse, i.e. this system is
simply non-constrained gravitationally, then according to the
conventional QFT $S_{\max}(\Re)\sim V(\Re)$, where $V(\Re)$ is the
bulk of $\Re$. However in Holographic Principle case, as it has
been demonstrated by G. 't Hooft and  R. V. Buniy and S. D. H. Hsu
in \cite{Hooft1},\cite{Hsu}
\begin{equation}\label{Hol9}
S_{\max}(\Re) \sim \frac{A(\Re)^{3/4}}{{\l_p}^2},
\end{equation}
And as was be shown in \cite{shalyt14} $QFT^{\alpha}$  will be
compatible with the holographic principle, i.e. with the
holographic entropy bound derived in (\ref{Hol9}). In terms of the
deformation parameter $\alpha$ the principal values of the Vacuum
Energy Problem may be simply and clearly defined. Let us begin
with the Schwarzschild black holes, whose
 semiclassical entropy is given by
\begin{equation}\label{D1}
S = \pi {R_{Sch}^2}/ l_p^2=\pi {R_{Sch}^2}
M_p^2=\pi\alpha_{R_{Sch}}^{-1},
\end{equation}
with the assumption that in the formula for $\alpha$ $R_{Sch}=x$ is
the measuring scale and $l_p = 1/M_p$. Here $R_{Sch}$
is the adequate Schwarzschild radius, and $\alpha_{R_{Sch}}$
is the value of $\alpha$ associated with this radius. Then, as it
has been pointed out in \cite{Bal}), in case the Fischler-
Susskind cosmic holographic conjecture \cite{Sussk1} is valid,
the entropy of the Universe is limited by its "surface"
 measured in Planck units \cite{Bal}:
\begin{equation}\label{D2}
S \leq \frac{A}{4} M_P^2,
\end{equation}
where the surface area $A = 4\pi R^2$ is defined in
terms of the apparent (Hubble) horizon
\begin{equation}\label{D3}
R = \frac{1}{\sqrt{H^2+k/a^2}},
\end{equation}
with curvature $k$  and scale $a$ factors.
\\ Again, interpreting $R$ from (\ref{D3}) as a measuring scale,
we directly obtain(\ref{D2}) in terms of $\alpha$:
\begin{equation}\label{D4}
S \leq \pi\alpha_{R}^{-1},
\end{equation}
where $\alpha_{R}=l^{2}_{p}/R^{2}$. Therefore, the average entropy
density may be found as
\begin{equation}\label{D5}
\frac{S}{V}\leq \frac{\pi \alpha_{R}^{-1}}{V}.
\end{equation}
Using further the reasoning line of \cite{Bal} based on
the results of the  holographic thermodynamics, we can relate
the entropy and energy of a
holographic system \cite{Jac1,Cai1}. Similarly, in terms of the
$\alpha$ parameter one can easily estimate the upper limit for
the energy density of the Universe (denoted here by $\rho_{hol}$):
\begin{equation}\label{D6}
\rho_{hol} \leq \frac{3}{8 \pi R^2} M_P^2 = \frac{3}{8
\pi}\alpha_{R} M_P^4,
\end{equation}
that is drastically differing from the one obtained with a naive
QFT
\begin{equation}\label{D7}
\rho^{naive}_{QFT}\sim M_P^4.
\end{equation}
Here by $\rho^{naive}_{QFT}$ we denote the energy Density
calculated from the naive QFT \cite{Zel1}. Obviously, as
$\alpha_{R}$ for $R$ determined by (\ref{D3}) is very small,
actually approximating zero, $\rho_{hol}$ is by several orders of
magnitude smaller than the value expected in QFT -
$\rho^{naive}_{QFT}$.
\\In fact, the upper limit of the right-hand side of(\ref{D6})
is attainable, as it has been demonstrated in \cite{Bou3} and
indicated in \cite{Bal}.
The "overestimation" value of $r$ for the energy
density $\rho^{naive}_{QFT}$, compared to
$\rho_{hol}$, may be determined as
\begin{equation}\label{D8}
r =\frac{\rho^{naive}_{QFT}}{\rho_{hol}}=\frac{8
\pi}{3}{\bf \alpha_{R}^{-1}}
 =\frac{8 \pi}{3} \frac{R^2}{L_P^2}
 =\frac{8 \pi}{3} \frac{S}{S_P},
\end{equation}
where $S_P$ is the entropy of the Plank mass and length
for the Schwarzschild black hole. It is clear that due to
smallness of $\alpha_{R}$ the value of $\alpha_{R}^{-1}$
is on the contrary too large. It may be easily calculated
(e.g., see \cite{Bal})
\begin{equation}\label{D9}
r = 5.44\times 10^{122}
\end{equation}
in a good agreement with the astrophysical data.
\\ Naturally, on the assumption that the vacuum energy density
$\rho_{vac}$ is involved in $\rho$ as a term
\begin{equation}\label{vac1}
\rho = \rho_M + \rho_{vac},
\end{equation}
where $\rho_M$ - average matter  density, in case of $\rho_{vac}$
we can arrive to the same upper limit (right-hand side of the
formula(\ref{D6})) as for $\rho$.
\section{Fischler-Susskind Conjecture and Gravitational Holography}
In this Section the arguments in support of the Fischler-Susskind
cosmic holographic conjecture are given on the basis of the results
obtained lately on Gravitational Holography.
\\ Quite recently, T.Padmanabhan in a series of his papers
\cite{Padm3}--\cite{Padm9} and some other works has convincingly
demonstrated that  Einstein equations may be derived from the
surface term of the GR Lagrangian, in fact containing the same
information as the bulk term.
\\And as Einstein-Hilbert's Lagrangian has the structure
$L_{EH}\propto R\sim (\partial g)^2+ {\partial^2g}$, in the usual
approach the surface term arising from  $L_{sur}\propto
\partial^2g$ has to be  canceled to get Einstein
equations from  $L_{bulk}\propto (\partial g)^2$ \cite{Padm10}.
But due to the relationship between $L_{bulk}$ and $L_{sur}$
\cite{Padm5}--\cite{Padm7},\cite{Padm10}, we have
 \begin{equation}
    \sqrt{-g}L_{sur}=-\partial_a\left(g_{ij}
\frac{\partial
\sqrt{-g}L_{bulk}}{\partial(\partial_ag_{ij})}\right)
\end{equation}
In such a manner one can suggest a holographic character of
gravity in that the bulk and surface terms of the gravitational
action contain identical information. However, there is a
significant difference between the first case when variation of
the metric $g_{ab}$ in $L_{\rm bulk}$ leads to Einstein equations,
and the second case associated with derivation of the GR field
equations from the action principle using only the surface term
and virtual displacements of horizons \cite{Padm2}, whereas the
metric is not treated as a dynamical variable \cite{Padm10}.
\\In the case under study, it is assumed from the beginning
that we consider the spaces with horizon. It should be noted that
in the Fischler-Susskind cosmic holographic conjecture it is
implied that the Universe represents spherically symmetric
space-time, on the one hand, and has a (Hubble) horizon
(\ref{D3})), on the other hand. But proceeding from the results of
\cite{Padm3}-- \cite{Padm10}, an entropy boundary is actually
given by the surface of horizon measured in Planck's units of area
\cite{Padm6}:
\begin{equation}\label{D10}
S=\frac{1}{4}\frac{A_H}{{{\l_p}^2}},
\end{equation}
where $A_H$ is the horizon area.
\\Because of this, it should be noted that Einstein's
equations may be obtained from the proportionality of the entropy
and horizon area together with the fundamental thermodynamic
relation connecting heat, entropy, and temperature \cite{Jac1}. In
fact \cite{Padm3}-- \cite{Padm10}, this approach has been extended
and complemented by the demonstration of holographicity  for the
gravitational action (see also \cite{Padm11}).
\\To sum it up, an assumption that space-time
is spherically symmetric and has a horizon is the only natural
assumption held in the Fischler-Susskind cosmic holographic
conjecture to support its validity. Then there is a resemblance to
thermodynamic systems, and one can associate the notions of
temperature and entropy with them. In the case of Einstein-Hilbert
gravity, it is possible to interpret Einstein's equations as the
thermodynamic identity $TdS = dE + PdV$ \cite{Padm12}.
\section{Notes}
I.This note is devoted to the demonstration of the fact, that in
case of the holographic principle validity in terms of the new
deformation parameter $\alpha$ in $QFT^{\alpha}$, considered above
and introduced  as early as 2002 \cite{shalyt11}--\cite{shalyt13},
all the principal values associated with the Vacuum (Dark) Energy
Problem may be defined simply and naturally. At the same time,
there is no place for such a parameter in the well-known QFT,
whereas in QFT with the fundamental length, specifically  in
$QFT^{\alpha}$ it is quite natural
\cite{shalyt1,shalyt2,shalyt4,shalyt6,shalyt7,shalyt9}.
\\
\\II. As indicated in \cite{shalyt14}, $QFT^{\alpha}$ (similar to
the conventional QFT) conforms to the Holographic Principle, being
coincident with QFT to a high accuracy in a semiclassical
approximation, i.e. for $\alpha\rightarrow 0$. In this case
$\alpha$ is small rather than vanishing. Specifically, the
smallness of $\alpha_{R}$ results in a very great value of $r$ in
(\ref{D8}),(\ref{D9}). Besides, from (\ref{D8}) it follows that
there exists some minimal entropy $S_{min} \sim S_P$, and this is
possible  only in QFT with the fundamental length.
\\
\\III.It should be noted that this section is related to Section 3
in \cite{Padm1} as well as to Sections 3 and 6 in \cite{Padm2}.
The constant $L_\Lambda$ introduced in these works is such that in
case under consideration $\Lambda\equiv l_\Lambda^{-2}$ is
equivalent to $R$, i.e. $\alpha_{R}\approx \alpha_{l_\Lambda}$
with $\alpha_{l_\Lambda}=l^{2}_{p} /l^{2}_\Lambda$. Then
expression in the right-hand side of (\ref{D6}) is the major term
of the formula for $\rho_{vac}$, and its quantum corrections are
nothing else as a series expansion in terms of
$\alpha_{l_\Lambda}$ (or $\alpha_{R}$):
\begin{equation}\label{D10}
\rho_{\rm vac}\sim
{\frac{1}{l_P^4}\left(\frac{l_P}{L_\Lambda}\right)^2}
+{\frac{1}{l_P^4}\left(\frac{l_P}{l_\Lambda}\right)^4} + \cdots
=\alpha_{l_\Lambda} M_P^4+\alpha^{2}_{l_\Lambda} M_P^4+...
\end{equation}
 In the first variant presented in \cite{Padm1} and
\cite{Padm2} the right-hand side (\ref{D10}) (formulas (12),(33)
in \cite{Padm1} and \cite{Padm2}, respectively)reveals an enormous
additional term $M_P^4\sim \rho^{naive}_{QFT}$ for
renormalization. As indicated in the previous Section, it may be,
however, ignored because the gravity is described by a pure
surface term. And in case under study, owing to the Holographic
Principle, we may proceed directly to (\ref{D10}). Moreover, in
$QFT^{\alpha}$ there is no need in renormalization as from the
start we are concerned with the ultraviolet-finiteness.

%References

\end{document}